\documentclass[12pt]{article}

\usepackage{amsfonts}
\usepackage{amssymb}
\usepackage{amscd}

\date{}
\begin{document}

 \title{\Large{\bf $p$-Adic and Adelic Quantum Mechanics\footnote{\textsf{\,
 Talk presented at the {\it First International Conference on $p$-Adic Mathematical Physics},
 1-5 October 2003, Moscow, Russia}}  }}

\author{Branko
Dragovich\,\footnote{\textsf{\, E-mail:\,dragovich@phy.bg.ac.yu}}\\
 {\it Institute of Physics, P.O. Box 57,} \\ {\it 11001
Belgrade, Serbia and Montenegro}}

\maketitle

\begin{abstract}

$p$-Adic mathematical physics emerged as a result of efforts to
find a non-Archimedean approach to the spacetime and string
dynamics at the Planck scale. One of its main achievements is a
successful formulation and development of $p$-adic and adelic
quantum mechanics, which have complex-valued wave functions of
$p$-adic and adelic arguments, respectively. Various aspects of
these quantum mechanics are reviewed here. In particular, the
corresponding Feynman's path integrals, some  minisuperspace
cosmological models, and relevant approache to string theory, are
presented. As a result of adelic approach, $p$-adic effects
exhibit a spacetime and some other discreteness, which depend on
the adelic quantum state of the physical system under
consideration. Besides review, this article contains also some new
results.
\end{abstract}

\section{Introduction}

 At the transition from 19th to 20th century   two great parts of
fundamental science were born: {\it Quantum Physics} and {\it
$p$-Adic Mathematics}. Developing quite independently till the
last two decades, they started to interact rather successfully so
that not only some quantum but also classical $p$-adic models have
been constructed and investigated.  As a result, in 1987 emerged
{\it $p$-Adic Mathematical Physics}, which is a basis to explore
various $p$-adic aspects of modern theoretical physics. Among the
main achievements of this new branch of contemporary mathematical
physics are {\it $p$-Adic and Adelic Quantum Mechanics}.

There are many physical and mathematical motivations to employ
$p$-adic numbers and adeles in investigation of mathematical and
theoretical aspects of modern quantum physics. Some primary of
them are: ({\it i}) the field of rational numbers ${\mathbb Q}$,
which contains all observational and experimental numerical data,
is a dense subfield not only in the field of real numbers
${\mathbb R}$ but also in the fields of $p$-adic numbers ${\mathbb
Q}_p$, ({\it ii}) there is a sufficiently well developed analysis
\cite{vladimirov1} within and over ${\mathbb Q}_p$ analogous to
that one related to ${\mathbb R}$, ({\it iii}) local-global
(Hasse-Minkowski) principle which states that usually when
something is valid on all local fields $({\mathbb R}, \, {\mathbb
Q}_p)$ is also valid on the global field $({\mathbb Q})$, ({\it
iv}) fundamental physical laws and relevant general mathematical
methods should be invariant \cite{volovich1} under an interchange
of the number fields ${\mathbb R}$ and ${\mathbb Q}_p$, ({\it v})
question "Which aspects of the Universe  cannot be described
without use of $p$-adic numbers ?" , ({\it vi}) there is a generic
quantum gravity uncertainty $\Delta x$ (see (\ref{1.1})) for
possible measurements of distances approaching to the Planck
length $\ell_0$, which restricts priority of Archimedean geometry
based on real numbers and gives rise to employment of the
non-Archimedean one related to $p$-adic numbers, ({\it vii}) it
seems to be quite reasonable to extend standard Feynman's  path
integral over real space to $p$-adic spaces, and ({\it viii})
adelic quantum mechanics \cite{dragovich2}, which is quantum
mechanics on an adelic space and contains standard as well as all
$p$-adic quantum mechanics, is consistent with all the above
assertions.

It is worth to explain in some details the above motivation ({\it
vi}). Namely, according to various considerations, which take
together standard quantum and gravitational principles, it follows
a strong generic restriction on experimental investigation of the
space-time structure at very short distances due to the relation
\begin{equation}
\Delta x \geq \ell_0 = \sqrt {\frac{\hbar G}{c^3}} \sim 10^{-33}
cm,     \label{1.1}
\end{equation}
where $\Delta x$ is an uncertainty measuring a distance, $\ell_0$
is the Planck length, $\hbar =\frac{h}{2\pi}$ is the reduced
Planck constant, $G$ is Newton's gravitational constant and $c$ is
the speed of light in vacuum. The uncertainty (\ref{1.1}) means
that one cannot measure distances smaller than $\ell_0$ and
$\ell_0$ can be regarded as a minimal (fundamental) length.
However, this result is derived assuming that whole space-time can
be described only by real numbers and Archimedean geometry. But we
cannot {\it a priory} exclude $p$-adic numbers and their
non-Archimedean geometric properties. To get a more complete
insight into structure of space-time at the Planck scale it is
quite natural to use adelic approach, which treats  simultaneously
and at an equal footing real (Archimedean) and $p$-adic
(non-Archimedean) aspects.

According to (\ref{1.1}), approach based only on Archimedean
geometry and real numbers predicts its own breakdown at the Planck
scale and gives rise to include $p$-adic non-Archimedean sector of
possible geometries. Namely, recall that having two segments on
straight line of different lengths $a$ and $b$, where $a<b$, one
can overpass the longer $b$ by applying the smaller $a$ some
$n$-times along $b$. In other words, if $a$ and $b$ are two
positive real numbers and $a<b$ then there exists an enough large
natural number $n$ such that $na>b$. This is an evident property
of the Euclidean spaces (and the field of real numbers), which is
known as Archimedean postulate, and can be extended to the
standard Riemannian spaces. One of the axioms of the metric spaces
is the triangle inequality which reads:
\begin{equation}
d(x,y) \leq d(x,z) + d(z,y),                  \label{1.2}
\end{equation}
where $d(x,y)$ is a distance  between points $x$ and $y$. However,
there is a subclass of metric spaces for which triangle inequality
is stronger in such way that:
\begin{equation}
d(x,y) \leq max \{d(x,z), d(z,y)  \}   \leq d(x,z) + d(z,y).
\label{1.3}
\end{equation}
Metric spaces with strong triangle inequality (\ref{1.3}) are
called non-Archimedean or ultrametric spaces. Since a measurement
means quantitative comparison of a given observable with respect
to a fixed value taken as its unit, it follows that a realization
of the Archimedean postulate is practically equivalent to the
measurements of distances. According to the uncertainty
(\ref{1.1}), it is not possible to handle distances shorter than
$10^{-33} cm$ and consequently one cannot apply  only Archimedean
geometry beyond the Planck length. Hence, for mathematical
modelling of space-time as well as matter (strings and branes)
when approaching to the Planck scale it is necessary to employ
adeles.

Before a systematic investigation of a possible adelic quantum
theory at the Planck scale it is useful to explore various aspects
of $p$-adic and adelic quantum mechanics.  These quantum mechanics
are well formulated and so far elaborated at the level which
promises their successful extension towards {\it Adelic
Superstring/M-theory}. This article contains a brief systematic
presentation of quantum mechanics on real, $p$-adic and adelic
spaces.

For a necessary  information on  usual properties of $p$-adic
numbers and related analysis one can see
\cite{vladimirov1,gouvea,schikhof,mahler,koblitz,gelfand}.

\section{Quantum mechanics on a real space}

This is ordinary (or standard) quantum mechanics (OQM). It has
four main sectors: Hilbert space, Quantization, Evolution and
Interpretation. Some of them can be formulated in a few different
ways, which are equivalent. Hilbert space of OQM consists of
square integrable complex-valued functions of real arguments,
which are mainly coordinates of $D$-dimensional space and time,
and is usually denoted by $L_2 ({\mathbb R}^D)$.

To physical opservables correspond linear self-adjoint operators
in $L_2 ({\mathbb R}^D)$. Classical dynamical expressions, which
depend on canonical variables $x_i, \, k_j$ of phase space, become
operators by quantization procedure usually initiated by the
Heisenberg algebra with commutation relations
\begin{equation} [\, \hat{x_i},\hat{k_j}\,] = i\,\hbar\,
\delta_{ij}, \qquad [\,\hat{x_i},\hat{x_j}\,] = 0,\qquad
[\,\hat{k_i},\hat{k_j}\,] =0,\label{2.1} \end{equation} where $i,
j=1,2,\dots ,D$. Note that instead of (\ref{2.1}) one can use an
equivalent quantization based on group relations $(h = 1)$
\begin{equation}
\chi_\infty (- \alpha_i {\hat x}_i) \, \chi_\infty (- \beta_j
\hat{k}_j) = \chi_\infty ( \alpha_i \beta_j \delta_{ij})\,
\chi_\infty (- \beta_j {\hat k}_j) \, \chi_\infty (- \alpha_i
\hat{x}_i), \label{2.2}
\end{equation}
\begin{equation}
\chi_\infty (- \alpha_i {\hat x}_i) \, \chi_\infty (- \alpha_j
\hat{x}_j) =  \chi_\infty (- \alpha_j {\hat x}_j) \, \chi_\infty
(- \alpha_i \hat{x}_i),    \label{2.3}
\end{equation}
\begin{equation}
\chi_\infty (- \beta_i {\hat k}_i) \, \chi_\infty (- \beta_j
\hat{k}_j) =  \chi_\infty (- \beta_j {\hat k}_j) \, \chi_\infty (-
\beta_i \hat{k}_i),    \label{2.4}
\end{equation}
where $\chi_\infty (u) =\exp (-2\pi i u)$ is real additive
character and $(\alpha_i, \beta_j)$ is a point of classical phase
space. Quantization of expressions which contain products of $x_i$
and $k_j$ is not unique. According to the Weyl quantization
\cite{weyl} any function $f (k,x)$, of classical canonical
variables $k$ and $x$, which has the Fourier transform
$\tilde{f}(\alpha, \beta)$ becomes a self-adjoint operator in $L_2
({\mathbb R}^D)$ in the following way:
\begin{equation}
{\hat f} ({\hat k}, {\hat x})= \int \chi_\infty (- \alpha {\hat x}
- \beta {\hat k})  {\tilde f}(\alpha, \beta) \, d^D\alpha \,
d^D\beta . \label{2.5}
\end{equation}

Evolution of the elements $\Psi (x,t)$ of $L_2 ({\mathbb R}^D)$ is
usually given by the Schr\"odinger equation \begin{equation}
\label{2.6} i\,\hbar\, \frac{\partial}{\partial t}\, \Psi (x,t) =
H ({\hat k},\, x)\,\Psi (x,t),      \end{equation} where $H$ is a
Hamiltonian and ${\hat k}_j = - i \hbar \frac{\partial}{\partial
x_j}$. Besides this differential equation there is also the
following integral form:
\begin{equation}
\psi (x'',t'') = \int {\cal K }_\infty(x'',t'';x',t')\, \psi
(x',t') \, d^Dx' , \label{2.7}
\end{equation}
where ${\cal K }_\infty(x'',t'';x',t')$ is the kernel of the
unitary representation of the evolution operator $U_\infty
(t'',t')$ and is postulated by Feynman to be a path integral
\cite{feynman}
\begin{equation}\label{2.8} {\cal K}_\infty (x'',t'';x',t') =\int_{(x',t')}^{(x'',t'')}
\chi_\infty ( - S[\,q\,])\, {\cal D}q ,
\end{equation}
where  functional $ S[\,q\,] = \int_{t'}^{t''} L(\dot{q},q, t)\,
dt$ is the action for a path $q(t)$ in the classical Lagrangian
$L(\dot{q},q,t)$, and $x''=q(t''), \ x'=q(t')$ are end points with
the notation $x =(x_1,x_2,\dots,x_D)$ and $q=(q_1,q_2,\dots,q_D)$.
The kernel ${\cal K}_\infty (x'',t'';x',t')$ is also known as the
probability amplitude for a quantum particle to pass from position
$x'$ at time  $t'$ to another point $x''$ at $t''$, and is closely
related to the quantum-mechanical propagator and Green's function.
The integral in (\ref{2.8}) has an intuitive meaning that a
quantum-mechanical particle may propagate from $x'$ to $x''$ using
infinitely many paths which connect these two points and that one
has to sum probability amplitudes over all of them. Thus the
Feynman path integral means a continual (functional) summation of
single transition amplitudes $\exp \left ( \frac{i}{\hbar}\,
S[\,q\,]\right ) $ over all possible continual paths $q(t)$
connecting $x'=q(t')$ and $x''=q(t'')$. In Feynman's formulation,
the path integral (\ref{2.8}) is the limit of an ordinary multiple
integral over $N$ variables $q_i = q(t_i)$ when $N\longrightarrow
\infty$. Namely, the time interval $t''-t'$ is divided into $N +1$
equal subintervals and integration is performed for every $q_i \in
(-\infty,+\infty)$ at fixed time $t_i$. We will consider Feynman's
path integral also in next sections and especially in Sec. 4.

Interpretation of OQM is related to the scalar products  of
complex-valued functions in $L_2 ({\mathbb R}^D)$ and will not be
discussed here, but can be found in standard books on quantum
mechanics including Ref. \cite{weyl}.

\section{Quantum mechanics on $p$-adic and adelic spaces}

It is remarkable that quantum mechanics on a real space can be
generalized on $p$-adic spaces for any prime  number $p$. However
there is not a unique way to perform generalization. As a result
there are two main approaches - with complex-valued and $p$-adic
valued elements of the Hilbert space on ${\mathbb Q}_p^D$.
Approach with $p$-adic valued wave functions has been introduced
by Vladimirov and Volovich \cite{vladimirov2a,vladimirov1} and
developed by Khrennikov \cite{khrennikov1,khrennikov2}. $p$-Adic
quantum mechanics with complex-valued wave functions  is more
suitable for connection with ordinary quantum mechanics, and in
the sequel we will refer only to this kind of quantum mechanics on
$p$-adic spaces.

Since wave functions are complex-valued, one cannot construct a
direct analogue of the Schr\"odinger equation (\ref{2.6}) with a
$p$-adic version of Heisenberg algebra (\ref{2.1}). According to
the Weyl approach to quantization, canonical noncommutativity in
$p$-adic case should be introduced by operators ($h=1$)
\begin{equation}
\hat Q_p(\alpha) \psi_p(x)=\chi_p(-\alpha x)\psi_p(x) , \, \quad
\hat K_p(\beta)\psi_p(x)=\psi_p(x+\beta)  \label{3.1}
\end{equation}
which satisfy
\begin{equation}
\hat Q_p(\alpha)\hat K_p(\beta)=\chi_p(\alpha\beta) \hat
K_p(\beta)\hat Q_p(\alpha) , \label{3.2}
\end{equation}
where $\chi_p (u) = \exp (2\pi i \{ u \}_p)$ is additive character
on the field ${\mathbb Q}_p$ and $\{ u \}_p$ is the fractional
part of $u \in {\mathbb Q}_p$ .

Let ${\hat x}$ and ${\hat k}$ be some operators of position $x$
and momentum $k$, respectively. Let us define operators $\chi_v
(\alpha {\hat x})$ and  $\chi_v (\beta {\hat k})$ by formulas
\begin{equation}
\chi_v (\alpha {\hat x}) \, \chi_v (a  x) = \chi_v (\alpha { x})
\, \chi_v (a { x}), \, \quad  \chi_v (\beta {\hat k})\, \chi_v (b
{ k}) = \chi_v (\beta { k})\, \chi_v (b k) ,  \label{3.3}
\end{equation}
where index $v$ denotes real $(v=\infty)$ and any $p$-adic case,
i.e. $v = \infty, 2,  \cdots, p, \cdots$ taking into account all
non-trivial and inequivalent valuations on ${\mathbb Q}$. It is
evident that these operators also act on a function $\psi_v (x)$ ,
which has the Fourier transform ${\tilde \psi}(k)$, in the
following way:
\begin{equation}
\chi_v (-\alpha {\hat x})\, \psi_v (x) = \chi_v (-\alpha {\hat
x})\, \int \chi_v (- k x)\, {\tilde \psi}(k)\, d^Dk =
\chi_v(-\alpha x)\, \psi_v(x) ,  \label{3.4}
\end{equation}
\begin{equation}
\chi_v (-\beta {\hat k})\, \psi_v (x) = \int \chi_v (-\beta k)\,
\chi_v (- k x)\, {\tilde \psi}(k)\, d^Dk =  \psi_v(x + \beta)  ,
\label{3.5}
\end{equation}
where integration in $p$-adic case is with respect to the Haar
measure $dk$ with the properties: $d(k +a) = dk,\, \,
d(ak)=|a|_p\, dk$ and $ \int_{|k|_p\leq 1} dk =1$. Comparing
(\ref{3.1}) with (\ref{3.4}) and (\ref{3.5}) we conclude that $
{\hat Q}_p(\alpha) = \chi_p(-\alpha {\hat x}) , \, {\hat
K}_p(\beta) = \chi_p(- \beta {\hat k})$. Now group relations
(\ref{2.2}), (\ref{2.3}), (\ref{2.4}) can be straightforwardly
generalized, including $p$-adic  cases, by replacing formally
index $\infty$ by $v$ . Thus we have
\begin{equation}
\chi_v (- \alpha_i {\hat x}_i) \, \chi_v (- \beta_j \hat{k}_j) =
\chi_v ( \alpha_i \beta_j \delta_{ij})\, \chi_v (- \beta_j {\hat
k}_j) \, \chi_v (- \alpha_i \hat{x}_i), \label{3.6}
\end{equation}
\begin{equation}
\chi_v (- \alpha_i {\hat x}_i) \, \chi_v (- \alpha_j \hat{x}_j) =
\chi_v (- \alpha_j {\hat x}_j) \, \chi_v (- \alpha_i \hat{x}_i),
\label{3.7}
\end{equation}
\begin{equation}
\chi_v (- \beta_i {\hat k}_i) \, \chi_v (- \beta_j \hat{k}_j) =
\chi_v (- \beta_j {\hat k}_j) \, \chi_v (- \beta_i \hat{k}_i).
\label{3.8}
\end{equation}

  One can introduce the unitary operator
\begin{equation}
W_v(\alpha {\hat x}, \beta {\hat k})=\chi_v(\frac{ 1}{ 2} \alpha
\beta)\, \chi_v (- \beta {\hat k}) \, \chi_v(- \alpha {\hat x}),
\label{3.9}
\end{equation}
which satisfies the Weyl relation
\begin{equation}
W_v(\alpha {\hat x}, \beta {\hat k})\, W_v(\alpha' {\hat x},
\beta' {\hat k}) =\chi_v(\frac{ 1}{ 2}( \alpha \beta' - \alpha'
\beta)) \, W_v((\alpha +\alpha'){\hat x}, (\beta +\beta') {\hat
k})
\label{3.10}
\end{equation}
and is a unitary representation of the Heisenberg-Weyl group.
Recall that this group consists of the elements $(z,\eta)$ with
the group product
\begin{equation} (z,\eta)\cdot (z',\eta
')=(z+z',\eta +\eta'+ \frac{1}{2} B(z,z')), \label{3.11}
\end{equation}
where $\quad z = (\alpha,\beta )\in {\mathbb Q}_v\times {\mathbb
Q}_v$  and $B(z,z') = \alpha \beta'- \beta \alpha'$ is a
skew-symmetric bilinear form on the phase space. Using operator
$W_v(\alpha {\hat x}, \beta {\hat k})$ one can generalize Weyl
formula for quantization (\ref{2.5}) and it reads
\begin{equation}
{\hat f}_v ({\hat k}, {\hat x})= \int W_v (\alpha {\hat x}, \beta
{\hat k}) \, {\tilde f}_v(\alpha, \beta) \, d^D\alpha \, d^D\beta
. \label{3.12}
\end{equation}

It is worth noting that equation (\ref{3.5}) suggests to introduce
\begin{equation}
\{ \beta {\hat k} \}_p^n \psi_p (x) = \int \{ \beta  k \}_p^n
\chi_p (-k\, x) {\tilde\psi}_p (k) \, d^Dk \label{3.13}
\end{equation}
which may be regarded as a new kind of the $p$-adic
pseudodifferential operator (for a successful  Vladimirov
pseododifferential operator, see \cite{vladimirov1}). Also
equation (\ref{3.6}) suggests a $p$-adic analogue of the
Heisenberg algebra in the form $(h=1)$
\begin{equation}
\{\alpha_i {\hat x_i}  \}_p \, \{\beta_j {\hat k_j} \}_p -
\{\beta_j {\hat k_j} \}_p \, \{\alpha_i {\hat x_i} \}_p =
-\frac{i}{2\pi}\, \delta_{ij}\, \{ \alpha \beta\}_p \, .
\label{3.14}
\end{equation}


As a basic instrument to treat dynamics of a $p$-adic quantum
model  is natural to take the kernel ${\cal K}_p(x'',t'';x',t')$
of the evolution operator $U_p(t'',t')$. This kernel obtains by
generalization of its real analogue starting from (\ref{2.7}) and
(\ref{2.8}), i.e.
\begin{equation}
\psi_v (x'',t'') = \int {\cal K }_v(x'',t'';x',t')\, \psi_v
(x',t') \, d^Dx' , \label{3.14}
\end{equation}
and
\begin{equation}\label{3.15} {\cal K}_v (x'',t'';x',t') =\int_{(x',t')}^{(x'',t'')}
\chi_v ( - \int_{t'}^{t''} L({\dot q}, q, t)\, dt)\, {\cal D}q .
\end{equation}

According to Vladimirov and Volovich
\cite{vladimirov2a,vladimirov2, vladimirov1}, $p$-adic quantum
mechanics  is given by a triple
\begin{equation}
(L_2({\mathbb Q}_p), W_p(z), U_p(t)), \label{3.16}
\end{equation}
where $W_p (z)$ corresponds to our $W_p (\alpha {\hat x}, \beta
{\hat k})$. A similar formulation is done in \cite{ruelle1}, where
evolution operator  for one-dimensional systems is presented by a
unitary representation of an Abelian subgroup of  $SL (2, {\mathbb
Q}_p) $ instead of the path integral for the kernel $ {\cal K}_p
(x'',t'';x',t')$  (see also \cite{meurice1}).

Adelic quantum mechanics \cite{dragovich2} is a natural
generalization of the above formulation of ordinary and $p$-adic
quantum mechanics. Recall that an adele $x$
\cite{gelfand,weil,platonov} is an infinite sequence
\begin{equation}
  x= (x_\infty, x_2, \cdots, x_p, \cdots),
  \label{3.17}
\end{equation}
where $x_\infty \in {\mathbb R}$ and $x_p \in {\mathbb Q}_p$ with
the restriction that for all but a finite set $\bf S$ of primes
$p$ one has  $x_p \in {\mathbb Z}_p = \{x_p \in {\mathbb Q}_p\, :
|x_p|_p \leq 1 \}$.  Componentwise addition and multiplication are
standard arithmetical  operations on the ring of adeles ${\cal
A}$, which can be defined as
\begin{equation}
 {\cal A} = \bigcup_{{\bf S}} {\cal A} ({\bf S}),
 \ \  {\cal A}({\bf S}) = {\mathbb R}\times \prod_{p\in {\bf S}} {\mathbb Q}_p
 \times \prod_{p\not\in {\bf S}} {\mathbb Z}_p.         \label{3.18}
\end{equation}
Rational numbers are naturally embedded in the space of adeles.
${\cal A}$ is a locally compact topological space.

There are  two kinds of analysis over topological ring of adeles
${\cal A}$, which are generalizations of the corresponding
analyses over $\mathbb R$ and  ${\mathbb Q}_p$. The first one is
related to the mapping ${\cal A}\to {\cal A}$ and the other one to
${\cal A}\to {\mathbb C}$. In complex-valued adelic analysis it is
worth mentioning  an additive character
\begin{equation}
 \chi (x) = \chi_\infty (x_\infty) \prod_p \chi_p (x_p),
 \label{3.19}
\end{equation}
a multiplicative character
\begin{equation}
  |x|^s = |x_\infty|_\infty^s \prod_p |x_p|_p^s, \ \ s\in {\mathbb C},
                                                    \label{3.20}
\end{equation}
and elementary functions of the form
\begin{equation}
 \varphi_{\bf S} (x) = \varphi_\infty (x_\infty) \prod_{p\in {\bf S}} \varphi_p (x_p)
 \prod_{p\not\in {\bf S}} \Omega (|x_p|_p),           \label{3.21}
\end{equation}
where $\varphi_\infty (x_\infty)$ is an infinitely differentiable
function on ${\mathbb R}$ and $|x_\infty |_\infty^n \varphi_\infty
(x_\infty) \to 0$ as $|x_\infty|_\infty \to \infty$ for any $n\in
\{0,1,2,\cdots  \}$,  $\varphi_p (x_p)$ are some locally constant
functions with compact support, and
\begin{equation}
\Omega (|x_p|_p) = \left\{  \begin{array}{ll}
                 1,   &   |x_p|_p \leq 1,  \\
                 0,   &   |x_p|_p > 1 .
                 \end{array}    \right.
                 \label{3.22}
\end{equation}
All finite linear combinations of elementary functions
(\ref{3.21}) make the set $S({\cal A})$ of the Schwartz-Bruhat
adelic functions. The Fourier transform of $\varphi (x)\in S({\cal
A})$, which maps $S(\cal A)$ onto $S({\cal A})$, is
\begin{equation}
 \tilde{\varphi}(y) = \int_{\cal A} \varphi (x)\chi (xy)dx,
 \label{3.23}
\end{equation}
where $\chi (xy)$ is defined by (\ref{3.19}) and $dx = dx_\infty
dx_2 dx_3 \cdots$ is the Haar measure on ${\cal A}$.

The Hilbert space  $L_2({\cal A})$ contains  complex-valued
functions of adelic argument  with the following scalar product
and norm:
$$
 (\psi_1,\psi_2) = \int_{\cal A} \bar{\psi}_1(x)\, \psi_2(x)\, dx ,
\quad
  ||\psi|| = (\psi,\, \psi)^{\frac{1}{2}} < \infty .
$$
 A basis of $L_2({\cal A})$ may be given by the set of
orthonormal eigefunctions in a spectral problem of the evolution
operator $U (t)$, where $t\in {\cal A}$. Such eigenfunctions have
the form
\begin{equation}
\psi_{{\bf S},\alpha} (x,t) = \psi_n^{(\infty)}(x_\infty,t_\infty)
 \prod_{p\in {\bf S}} \psi_{\alpha_p}^{(p)} (x_p,t_p)
 \prod_{p\not\in {\bf S}} \Omega (|x_p|_p),
 \label{3.24}
\end{equation}
where $\psi_n^{(\infty)} \in L_2({\mathbb R})$ and
$\psi_{\alpha_p}^{(p)} \in L_2({\mathbb Q}_p) $ are eigenfunctions
in ordinary and $p$-adic cases, respectively. Indices $n,
\alpha_2, \cdots, \alpha_p, \cdots$ are related to the
corresponding real and $p$-adic eigenvalues of the same observable
in a physical system. $\Omega (|x_p|_p)$ is an element of
$L_2({\mathbb Q}_p)$, defined by (\ref{3.22}), which is invariant
under transformation of an evolution operator $U_p(t_p)$ and
provides convergence of the infinite product (\ref{3.24}). For a
fixed ${\bf S}$, states $\psi_{{\bf S},\alpha} (x,t)$ in
(\ref{3.24}) are eigefunctions of $L_2({\cal A}({\bf S}))$, where
${\cal A}({\bf S})$ is a subset of adeles ${\cal A}$ defined by
(\ref{3.18}). Elements of $L_2({\cal A})$ may be regarded as
superpositions $\psi (x) = \sum_{{\bf S},\alpha} C({\bf S},
\alpha)\, \psi_{{\bf S}, \alpha}(x), $ where $\psi_{{\bf
S},\alpha}(x)\in L_2({\cal A}({\bf S}))$  (\ref{3.24})  and
$\sum_{{\bf S},\alpha} |C({\bf S},\alpha)|_\infty^2 =1$.

Theory of $p$-adic generalized functions is presented in Ref.
\cite{vladimirov1}. There is not yet a theory of generalized
functions on adelic spaces, but there is some progress within
adelic quantum mechanics \cite{dragovich1a}  (see also
\cite{radyno}).

Adelic evolution operator $U(t)$ is defined by
\begin{equation} \nonumber
U(t'')\, \psi(x'')=\int_{{\cal A}} {\cal K}(x'',t'';x',t')\,
\psi(x',t')\, dx'
\end{equation}
\begin{equation}
=\prod\limits_{v}{} \int_{{\mathbb Q}_{v}}{\cal K}_v
(x''_{v},t''_v;x'_v,t'_v)\, \psi_v(x'_v,t'_v)\, dx'_v.
\label{3.25}
\end{equation}
where $v=\infty, 2, 3,\cdots, p,\cdots$. The eigenvalue problem
for $U(t)$ reads
\begin{equation} U(t)\, \psi _{{\bf S},\alpha } (x)=\chi
(E_{\alpha}\, t)\,  \psi _{{\bf S},\alpha} (x) , \label{3.26}
\end{equation}
where $\psi_{{\bf S},\alpha } (x)$ are adelic eigenfunctions
(\ref{3.24}), and $E_{\alpha }=(E_{\infty}, E_{2},..., E_{p},...)$
is the corresponding adelic energy.

Adelic quantum mechanics takes into account ordinary as well as
$p$-adic quantum effects and may be regarded as a starting point
for construction of a more complete quantum cosmology, quantum
field theory and string/M-theory. In the  limit of large distances
adelic quantum mechanics effectively becomes the ordinary one
\cite{dragovich6}.

\bigskip

\section{$p$-Adic and adelic path integrals, and some simple quantum models}
\bigskip

$p$-Adic path integral was initiated in \cite{vladimirov2} and by
subdivision of the time interval was computed for the harmonic
oscillator \cite{zelenov} and for a particle in a constant field
\cite{dragovich5}. Analytic evaluation of path integral for
quantum-mechanical systems with general form of quadratic
Lagrangians in the same way for real and $p$-adic cases is
performed in \cite{dragovich3,dragovich4}.

Starting from (\ref{3.14}) one can easily derive the following
three general properties:
\begin{equation}
  \int{\cal K}_v(x'',t'';x,t) {\cal K}_v(x,t;x',t') dx =
{\cal K}_v(x'',t'';x',t') ,      \label{4.1}
\end{equation}
\begin{equation}
   \int \bar{{\cal K}}_v(x'',t'';x',t') {\cal K}_v(y,t'';x',t') dx' =
\delta_v (x''-y) ,       \label{4.2}
\end{equation}
\begin{equation}
   {\cal K}_v(x'',t'';x',t'') = \lim_{t'\to t''} {\cal K}_v(x'',t'';x',t') =
\delta_v (x''-x') .    \label{4.3}
\end{equation}

Quantum fluctuations lead to deformations of classical trajectory
and any quantum history may be presented as $q(t) = x(t) + y(t)$,
where $y'=y(t')=0$  and  $y''=y(t'')=0$. For Lagrangians
$L(\dot{q},q,t)$ which are quadratic polynomials in ${\dot q}$ and
$q$,  the corresponding Taylor expansion of $S[q]$ around
classical path $x(t)$  is
\begin{equation}
   S[q] = S[x] + \frac{1}{2!} \delta^2 S[x] = S[x] +
   \frac{1}{2}\int_{t'}^{t''} \left( \dot{y}\frac{\partial}{\partial \dot{q}} +
   y\frac{\partial}{\partial q}   \right)^2 L(\dot{q},q,t)dt ,
   \label{4.4}
\end{equation}
where we used $\delta S[x] = 0$. Hence we get
 $$   
   {\cal K}_v (x'',t'';x',t') = \chi_v \left( -\frac{1}{h}
   \bar{S}(x'',t'';x',t') \right)
$$ 
\begin{equation}
 \times\int_{(y'\to 0,t')}^{(y''\to 0,t'')}
   \chi_v \left(- \frac{1}{2h} \int_{t'}^{t''} \left( \dot{y}
   \frac{\partial}{\partial\dot{q}} +
    y \frac{\partial}{\partial q} \right)^2 L(\dot{q},q,t)dt
   \right) {\cal D}y,
    \label{4.5}
\end{equation}
where  $\bar{S}(x'',t'';x',t') = S[x]$.

From (\ref{4.5}) follows that ${\cal K}_v(x'',t'';x',t')$  has the
form
\begin{equation}
    {\cal K}_v(x'',t'';x',t') = N_v(t'',t') \chi_v \left( -\frac{1}{h}
    \bar{S}(x'',t'';x',t')\right) ,   \label{4.6}
\end{equation}
where $N_v(t'',t')$ does not depend on end points $x''$ and  $x'$.

To calculate $N_v(t'',t')$ one can use (\ref{4.1}) and (\ref{4.2})
(see, e.g. \cite{dragovich4}). Then one obtains
 that $v$-adic kernel ${\cal K}_v(x'',t'';x',t')$ of the unitary
evolution operator for one-dimensional systems with quadratic
Lagrangians has the form
$$
{\cal K}_v(x'',t'';x',t') = \lambda_v\left( -\frac{1}{2h}
\frac{\partial^2}{\partial x'' \partial x'}\bar{S}(x'',t'';x',t')
\right)
$$
\begin{equation}
\times\left\vert \frac{1}{h} \frac{\partial^2} {\partial x''
\partial x'}\bar{S}(x'',t'';x',t') \right\vert_v^{\frac{1}{2}}
\chi_v\left( -\frac{1}{h} \bar{S}(x'',t'';x',t')  \right),
                                                      \label{4.7}
\end{equation}
where $\lambda_v$-functions are defined in \cite{vladimirov1}.

For practical considerations, we define adelic path integral in
the form
\begin{equation}
  {\cal K}_{\cal A}(x'',t'';x',t') = \prod_v
\int_{(x_v',t_v')}^{(x_v'',t_v'')} \chi_v \left( -\frac{1}{h}
\int_{t_v'}^{t_v''} L (\dot{q}_v,q_v,t_v) dt_v  \right) {\cal
D}q_v . \label{4.8}
\end{equation}
 As an adelic Lagrangian one understands an infinite sequence
\begin{equation}
 L_{\cal A}(\dot{q},q,t) = (L(\dot{q}_\infty,q_\infty,t_\infty),
L(\dot{q}_2,q_2,t_2), L(\dot{q}_3,q_3,t_3),\cdots,
 L(\dot{q}_p,q_p,t_p),\cdots),
 \label{4.9}
\end{equation}
where $|L(\dot{q}_p,q_p,t_p)|_p \leq 1$ for all primes $p$ but a
finite set $\bf S$ of them.

When one has a system with more than one dimension and coordinates
are uncoupled, then the total $v$-adic path integral is  product
of the corresponding one-dimensional path integrals. Investigation
of the coupled case is in a progress.

As an illustration of $p$-adic and adelic quantum-mechanical
models, the following one-dimensional systems with the quadratic
Lagrangians were considered: 1) $L({\dot q}, q) = \frac{m}{2}\,
{\dot q}^2$, a free particle  \cite{vladimirov1,dragovich2} , 2)
$L({\dot q}, q) = \frac{m}{2}\, {\dot q}^2 + a q$, a particle in a
constant field \cite{dragovich5}, 3) $L({\dot q}, q) =
\frac{m}{2}\, {\dot q}^2 \, - \frac{m\, \omega^2 }{2}\, q^2$, a
harmonic oscillator \cite{vladimirov1,dragovich2}, 4) $L({\dot q},
q) = -m\, c^2 \, \sqrt{{\dot q}_\mu \, {\dot q}^\mu} $, a free
relativistic particle \cite{dragovich6} and 5) $L({\dot q}, q) =
\frac{m}{2}\, {\dot q}^2 \, - \frac{m\, \omega^2 (t)}{2}\, q^2$, a
harmonic oscillator with time-dependent frequency
\cite{dragovich7}.

Let us mention that when time is real and trajectories are
$p$-adic , and vice versa, possible  functional integrals are
considered by Parisi \cite{parisi}. There is another proposal for
a path integral formulation of some evolution operators with
$p$-adic time \cite{meurice2}. For an approach to adelic path
space integrals with real time, see \cite{blair}.

\bigskip
\section{$p$-Adic and adelic quantum cosmology}

The main task of quantum cosmology is to describe  the very early
stage in the evolution of the Universe. At this stage, the
Universe was in a quantum state, which should be described by a
wave function. Usually one takes  that this wave function is
complex-valued and depends on some real parameters. Since quantum
cosmology is related to the Planck scale phenomena it is natural
to reconsider its foundations. We  maintain here the standard
point of view that the wave function takes complex values, but we
treat its arguments (space-time coordinates, gravitational and
matter fields) to be not only real but also $p$-adic and adelic.

There is not $p$-adic generalization of the Wheeler - De Witt
equation for cosmological models. Instead of differential
approach, Feynman's path integral method was exploited
\cite{dragovich8} and minisuperspace cosmological models are
investigated as models of adelic quantum mechanics
\cite{dragovich9,dragovich10}.

$p$-Adic and  adelic minisuperspace quantum cosmology is an
application of $p$-adic and adelic quantum mechanics to the
cosmological models, respectively. In the path integral approach
to standard quantum cosmology, the starting point is Feynman's
path integral method, {\it i.e.} the amplitude to go from one
state with intrinsic metric $h_{ij}'$ and matter configuration
$\phi'$ on an initial hypersurface $\Sigma'$ to another state with
metric $h_{ij}''$ and matter configuration $\phi''$ on a final
hypersurface $\Sigma''$ is given by the path integral
\begin{equation}
{\cal K}_\infty ( h_{ij}'',\phi'',\Sigma''; h_{ij}',\phi',\Sigma')
= \int  \chi_\infty(-S_\infty[g_{\mu\nu},\Phi])\, \, {\cal
D}{(g_{\mu\nu})}_\infty \, {\cal D}(\Phi)_\infty  \label{5.1}
\end{equation}
over all four-geometries $g_{\mu\nu}$ and matter configurations
$\Phi$, which interpolate between the initial and final
configurations. In (\ref{5.1}) $S_\infty [g_{\mu\nu},\Phi]$ is an
Einstein-Hilbert action for the gravitational and matter fields.
This action can be calculated using metric in the standard 3+1
decomposition
\begin{equation}
ds^2=g_{\mu\nu}dx^\mu dx^\nu=-(N^2 -N_i N^i)dt^2 + 2N_i dx^i dt +
h_{ij} dx^i dx^j, \label{5.2}
\end{equation}
where $N$ and $N_i$ are the lapse and shift functions,
respectively. To perform $p$-adic and adelic generalization we
make first $p$-adic counterpart of the action using
form-invariance under change of real to the $p$-adic number
fields. Then we generalize (\ref{5.1}) and introduce $p$-adic
complex-valued cosmological amplitude
\begin{equation}
 {\cal K}_p ( h_{ij}'',\phi'',\Sigma''; h_{ij}',\phi',\Sigma') =
 \int
\chi_p(-S_p[g_{\mu\nu},\Phi]) \, \, {\cal D}{(g_{\mu\nu})}_p \,
{\cal D}(\Phi)_p. \label{5.3}
\end{equation}

Since the space of all three-metrics and matter field
configurations on a three-surface, called superspace,  has
infinitely many dimensions, in computation one takes an
approximation. A useful approximation is to truncate the infinite
degrees of freedom to a finite number $q_\alpha(t)$,
($\alpha=1,2,...,n$). In this way, one obtains a  minisuperspace
model. Usually, one restricts the four-metric to be of the form
(\ref{5.2}), with $N^i=0$ and $h_{ij}$ approximated by
$q_\alpha(t)$. For the homogeneous and isotropic cosmologies  the
metric is a Robertson-Walker one, which spatial sector reads
\begin{equation}
h_{ij}dx^idx^j=a^2(t)\, d\Omega_3^2 = a^2(t) \left[
d\chi^2+\sin^2\chi(d\theta^2+\sin^2\theta d\varphi^2) \right],
\label{5.4}
\end{equation}
where $a(t)$ is a scale factor. If we use also a single scalar
field $\phi$, as a matter content of the model, minisuperspace
coordinates are $ a $ and $\phi$. Models can be also homogeneous
and anisotropic.

For the boundary condition $q_\alpha(t'')=q_\alpha''$, \
$q_\alpha(t')=q_\alpha'$ in the gauge $ N=1$, we have $v$-adic
minisuperspace propagator
\begin{equation}
 {\cal K}_v (q_{\alpha}''|q_{\alpha}') =\int dt \ {\cal
K}_v (q_{\alpha}'',t'';q_{\alpha}',t'), \quad t=t''-t' ,
\label{5.5}
\end{equation}
where
\begin{equation}
 {\cal K}_v (q_{\alpha}'',t'';q_{\alpha}',t') =\int
 \chi_v(-S_v[q_\alpha]) \, \, {\cal D}q_\alpha  \label{5.6}
\end{equation}
is an ordinary quantum-mechanical propagator between fixed
minisuperspace coordinates ($q_\alpha',q_\alpha''$)  in  fixed
times. $S_v$ is the $v$-adic  valued action of the minisuperspace
model which has the form
\begin{equation}
S_v[q_\alpha]= \int_{t'}^{t''} dt  \left[ \frac{1}{2}
f_{\alpha\beta}(q)\dot q^\alpha\dot q^\beta-U(q) \right],
\label{5.7}
\end{equation}
where $f_{\alpha\beta}$ is a minisuperspace metric
$(ds^2_m=f_{\alpha\beta}dq^\alpha dq^\beta)$ with a signature
($-,+,+,\dots$). This metric includes  gravitational and
 matter degrees of freedom.

The standard minisuperspace ground-state wave function in the
Hartle-Hawking (no-boundary) proposal \cite{hartle} is defined by
functional integration in the Euclidean version of
\begin{equation}
\psi_\infty[h_{ij}]= \int \chi_\infty(-S_\infty[g_{\mu\nu},\Phi])
\, \, {\cal D}(g_{\mu\nu})_\infty \, {\cal D}(\Phi)_\infty,
\label{5.8}
\end{equation}
over all compact four-geometries $g_{\mu\nu}$ which induce
$h_{ij}$ at the compact three-manifold. This three-manifold is the
only boundary of the all four-manifolds. Extending Hartle-Hawking
proposal to the $p$-adic minisuperspace  \cite{arefeva},  an
adelic Hartle-Hawking wave function is the infinite product
\begin{equation}
\psi[h_{ij}]= \prod_{v}\int  \chi_\upsilon(-S_v [g_{\mu\nu},\Phi])
\, \, {\cal D}(g_{\mu\nu})_v \, {\cal D}(\Phi)_v , \label{5.9}
\end{equation}
where path integration must be performed over both, Archimedean
and non-Archimedean geometries. When  evaluation of the
corresponding functional integrals for a minisuperspace model
yields $\psi(q_\alpha)$ in the form (\ref{3.24}), then we  say
that such cosmological model is a Hartle-Hawking adelic one. It is
shown \cite{dragovich8} that the de Sitter minisuperspace model in
$D=4$ space-time dimensions is  the Hartle-Hawking adelic one.

It is shown in \cite{dragovich9,dragovich10}   that $p$-adic and
adelic generalization of the minisuperspace cosmological models
can be successfully performed in the framework of $p$-adic and
adelic quantum mechanics \cite{dragovich2} without use of the
Hartle-Hawking approach.  The following cosmological models are
investigated: the de Sitter model, model with a homogeneous scalar
field, anisotropic Bianchi model with three scale factors and some
two-dimensional minisuperspace models. As a result of $p$-adic
effects and adelic approach, in these models there is some
discreteness of minisuperspace and cosmological constant. This
kind of discreteness was obtained for the first time in the
context of  the Hartle-Hawking adelic de Sitter quantum model
\cite{dragovich8}.

\bigskip

\section{Towards adelic string theory}

A notion of $p$-adic string was introduced in \cite{volovich2},
where the hypothesis on the existence of non-Archimedean geometry
at the Planck scale was made, and string theory with $p$-adic
numbers was initiated. In particular, generalization of the usual
Veneziano and Virasoro-Shapiro amplitudes with complex-valued
multiplicative characters over various number fields was proposed
and  $p$-adic valued Veneziano amplitude was constructed by means
of $p$-adic interpolation. Very successful  $p$-adic analogues of
the Veneziano and Virasoro-Shapiro amplitudes were proposed in
\cite{freund1} as the corresponding Gel'fand-Graev \cite{gelfand}
beta functions. Using this approach, Freund and Witten obtained
\cite{freund2} an attractive adelic formula
\begin{equation}
 A_\infty (a,b) \prod_p A_p (a,b) =1 ,            \label{6.1}
\end{equation}
which states that the product of the crossing symmetric Veneziano
(or Virasoro-Shapiro) amplitude and its all $p$-adic counterparts
equals unit (or a definite constant). This gives possibility to
consider an ordinary four-point function, which is rather
complicate, as an infinite product of its inverse $p$-adic
analogues, which have simpler forms. These first papers induced an
interest in various aspects of $p$-adic string theory (for a
review, see \cite{freund3,vladimirov1}). A recent interest in
$p$-adic string theory has been mainly related to  the tachyon
condensation \cite{sen}, nonlinear dynamics \cite{vladimirov4} and
an extension of $p$-adic and adelic path integrals to string
amplitudes  \cite{dragovich11}.

Like in the ordinary string theory, the starting point in $p$-adic
string theory is a construction of the corresponding scattering
amplitudes.  Recall that the ordinary crossing symmetric Veneziano
amplitude can be presented in the following four forms:
$A_\infty(k_1,\cdots,k_4) \equiv   A_\infty(a,b)$
\begin{equation}
= g^2 \int_{{\mathbb R}} \vert x \vert_\infty^{a-1}
  \vert 1-x\vert_\infty^{b-1} dx
     \label{6.2}
\end{equation}
\begin{equation}
 = g^2 \left[\frac{\Gamma{(a)} \Gamma{(b)} }{\Gamma{(a+b)}} +
 \frac{\Gamma{(b)}\Gamma{(c)}}{\Gamma{(b+c)}} +
 \frac{\Gamma{(c)}\Gamma{(a)}}{\Gamma{(c+a)}}\right]     \label{6.3}
\end{equation}
\begin{equation}
 = g^2 \frac{\zeta(1-a)}{\zeta(a)} \frac{\zeta(1-b)}{\zeta(b)}
 \frac{\zeta(1-c)}{\zeta(c)}                     \label{6.4}
\end{equation}
\begin{equation}
=g^2 \int {\cal D}X \exp\left(  -\frac{i}{2\pi}
 \int d^2 \sigma \partial^\alpha X_\mu \partial_\alpha X^\mu \right)
 \prod_{j=1}^4 \int d^2 \sigma_j \exp\left( i k_\mu^{(j)} X^\mu
\right) ,                           \label{6.5}
\end{equation}
where $\hbar=1,\ T=1/\pi$, and $a=-\alpha (s) = - 1 -\frac{s}{2},
\ b=-\alpha (t), \ c=-\alpha (u)$ with the condition $s+t+u = -8$,
i.e. $a+b+c=1$.

To introduce a $p$-adic Veneziano amplitude one can consider a
$p$-adic analogue of any of  the above four expressions. $p$-Adic
generalization of the first expression was proposed in
\cite{freund1} and it reads
\begin{equation}
 A_p (a,b) = g_p^2 \int_{{\mathbb Q}_p} \vert x\vert_p^{a-1}
 \vert 1-x\vert_p^{b-1} dx ,                 \label{6.6}
\end{equation}
where $\vert \cdot \vert_p$ denotes $p$-adic absolute value.
 In this case only
string world-sheet parameter $x$ is treated as $p$-adic variable,
and all other quantities maintain their usual (real) valuation. An
attractive adelic formula (\ref{6.1}) was found \cite{freund2}. A
similar product formula holds also for the Virasoro-Shapiro
amplitude. These infinite products are divergent, but they can be
successfully regularized. Unfortunately, there is a problem to
extend this formula to the higher-point functions.

$p$-Adic analogues of (\ref{6.2}) and (\ref{6.3}) were also
proposed in \cite{volovich2} and \cite{dragovich12}, respectively.
In these cases, world-sheet, string momenta and amplitudes are
manifestly $p$-adic. Since string amplitudes are $p$-adic valued
functions, it is not so far enough clear their physical
interpretation.

Expression (\ref{6.4}) is based on Feynman's path integral method,
which is generic for all quantum systems and has successful
$p$-adic generalization.  $p$-Adic counterpart of (\ref{6.4}) is
proposed in \cite{dragovich11} and has been partially elaborated
in \cite{dragovich13} and \cite{dragovich14}.  Note that in this
approach, $p$-adic string amplitude is complex-valued, while not
only the world-sheet parameters but also target space coordinates
and string momenta are $p$-adic variables. Such $p$-adic
generalization is a natural extension of the formalism of $p$-adic
\cite{vladimirov2} and adelic \cite{dragovich2} quantum mechanics
to string theory. This is a promising subject and should be
investigated in detail, and applied to the branes and M-theory,
which is presently the best candidate for the fundamental physical
theory at the Planck scale.

\bigskip

\section{Concluding remarks}

Among the very interesting and fruitful recent developments  have
been noncommutative geometry and  noncommutative field theory,
which  may be regarded as a deformation of the ordinary one in
which standard field multiplication is replaced by the Moyal
(star) product
\begin{equation}
 (f\star g)(x) =\exp\left[\frac{i \hbar}{2} \theta^{ij}\frac{
\partial}{\partial y^{i}} \frac{
\partial}{\partial z^j}\right] f(y)g(z)\vert_{y=z =x},
\label{7.1}
\end{equation}
where $x= (x^1,x^2,\cdots,x^d)$ is a spatial point  , and
$\theta^{ij} =-\theta^{ji}$ are noncommutativity parameters.
Replacing the ordinary product between noncommutative coordinates
by the Moyal product (\ref{7.1}) one obtains
\begin{equation}
 x^{i}\star x^j - x^j \star x^{i} = i\hbar \theta^{ij} .  \label{7.2}
\end{equation}
It is worth noting that one can introduce \cite{dragovich14} the
Moyal product in $p$-adic quantum mechanics and it reads
\begin{equation}
( f \ast  g)(x)=\int_{{\mathbb Q}_p^D}\int_{{\mathbb Q}_p^D} d^Dk
d^Dk' \ \chi_p(-(x^ik_i+x^jk'_j)+\frac{1}{2} k_ik'_j\,
\theta^{ij})\tilde f(k)\tilde g(k'), \label{7.3}
\end{equation}
where $D$ denotes  spatial dimensionality, and $\tilde f(k)$, \
$\tilde g(k')$ denote the Fourier transforms of $f(x)$ and $g(x)$.
Some real, $p$-adic and adelic aspects of the noncommutative
scalar solitons  are investigated in Ref. \cite{dragovich15}.

An  extension of the above formalism with Feynman's path integral
 to $p$-adic and adelic quantum field theory is  considered in
 \cite{dragovich16}. For a review on non-Archimedean Geometry and
 Physics on Adelic Spaces, see \cite{dragovich17}. Some remarks on
 arithmetic quantum physics are presented in \cite{varadarajan}.

\bigskip

\bigskip

\noindent {\bf Acknowledgements\, } The work on this paper was
supported in part by the Serbian Ministry of Science, Technologies
and Development under contract No 1426 and by RFFI grant
02--01-01084 .

\end{document}